# Two-fold superstructure of titanium(III)-oxybromide at $T = 17.5$ K

Lukš Palatinus,† Andreas Schönleber and Sander van Smaalen

*Laboratory for Crystallography, University of Bayreuth, D-95440 Bayreuth, Germany. E-mail: smash@uni-bayreuth.de*

**Abstract**

The low-temperature ($T$ = 17.5 K) structure of titanium(III)-oxybromide, TiOBr, is reported as a two-fold superstructure of the crystal structure at room temperature. Weak superlattice reflections could be measured with synchrotron radiation x-rays, and they were analysed by structure refinements employing superspace techniques. Both the low temperature and room temperature structures of TiOBr are isostructural to the corresponding structures of TiOCl. The results indicate that at low temperatures TiOBr is in a spin–Peierls state similar to TiOCl, but with the modulations and relevant interactions smaller than in the latter compound.

---

† Permanent address: Institute of Physics, Academy of Sciences of the Czech Republic, 16253 Prague 6, Czech Republic

**Comment**

TiOBr (von Schnering *et al.*, 1975) and TiOCl (Schäfer *et al.*, 1958) are isostructural compounds that crystallize in the FeOCl structure type (Fig. 1). On cooling they undergo two phase transitions, as evidenced by anomalies in the temperature dependencies of the magnetic susceptibilites (Seidel et al., 2003; Kato *et al.*, 2005; Lemmens *et al.*, 2005). Previously we have shown that in the lowest-temperature phase (T < 67 K) the crystal structure of TiOCl is a two-fold superstructure of the structure at room temperature (Shaz *et al.*, 2005). The superstructure involves the formation of Ti—Ti dimers on the chains of Ti atoms parallel to [010], thus suggesting a spin–Peierls state by direct exchange on the quasi-one-dimensional (1D) chains of Ti atoms (Shaz *et al.*, 2005). Recently Sasaki *et al.* (2005) reported superlattice reflections for TiOBr at low temperatures, but they did not report a satisfactory structure model, due to the limited number of measured reflections (2 reflections).

Here we report the low-temperature, two-fold superstructure of TiOBr at T = 17.5 K. It is found that the pattern of displacements in TiOBr is similar to that in TiOCl, but with amplitudes that are only about half of the displacement amplitudes in TiOCl (Fig. 2 and Table 1). These results indicate that TiOBr is in a spin–Peierls state at low temperatures similar to TiOCl, but with the relevant interactions smaller than in TiOCl, in accordance with the lower transition temperatures in TiOBr ($T_{c1}$ = 27 K; $T_{c2}$ = 47 K) than in TiOCl ($T_{c1}$ = 67 K; $T_{c2}$ = 90 K (Seidel *et al.*, 2003; Shaz *et al.*, 2005).

## Experimental

Single crystals for TiOBr were grown by gas transport following published procedures (Schäfer *et al.*, 1958). Starting materials for synthesis were Ti (Alpha, 99.99% purity), $TiO_2$ (Alpha, 99.99%) and $TiBr_4$ (Aldrich, 99.99%). They were mixed with a 40% surplus of Ti and $TiBr_4$. The sealed, evacuated ($p = 1.5\ 10^{-2}$ hPa) quartz glass tube was heated in a temperature gradient of 650:550 °C (923:823 K) for 72 h. The tube was cooled to room-temperature at a rate of 15.6 K $h^{-1}$. The yellowish-brown crystals have a needle to plate-like habitus and develop a greyish-white patina in air. Therefore they were further handled under inert gasses.

*Crystal data*

TiOBr
$M_r = 143.8$
Monoclinic
$P2_1/m\ 1\ 1$
$a = 3.7852\ (12)$ Å
$b = 6.9366\ (9)$ Å
$c = 8.501\ (3)$ Å
$\alpha = 90.00\ (2)$ °
$V = 223.21\ (10)$ Å$^3$
$Z = 4$
$D_x = 4.2772$ Mg m$^{-3}$
$D_m$ not measured

synchrotron radiation
$\lambda = 0.5$ Å
Cell parameters from 24 reflections
$\theta = 10.040$–$23.014°$
$\mu = 8.354$ mm$^{-1}$
$T = 17.5$ K
Platelet
Translucent light brown
0.27 × 0.13 × 0.002 mm

*Data collection*

Huber four-circle diffractometer
Profile data with $\omega$-scans
Absorption correction:
gaussian Jana2000 (Petricek *et al.*, 2000)
$T_{\min} = 0.4746$, $T_{\max} = 0.9776$
989 measured reflections
416 independent reflections
336 reflections with
$I > 3\sigma(I)$
$R_{\rm int} = 0.0542$

$\theta_{\max} = 20.48°$
$h = -5 \rightarrow 5$
$k = -9 \rightarrow 3$
$l = -10 \rightarrow 0$
3 standard reflections
frequency: 90 min
intensity decay: none

*Refinement*

Refinement on $F$

$R = 0.0182$

$wR = 0.0287$

$S = 1.03$

416 reflections

21 parameters

$w = 1/(\sigma^2(F) + 0.0004F^2)$

$(\Delta/\sigma)_{max} = 0.0001$

$\Delta\rho_{max} = 0.53$ e Å$^{-3}$

$\Delta\rho_{min} = -0.57$ e Å$^{-3}$

Extinction correction: $B$—$C$ type 1 Gaussian isotropic (Becker & Coppens, 1974)

Extinction coefficient: 0.09 (2)

Scattering factors from International Tables Vol C Tables 4.2.6.8 and 6.1.1.1

Table 1. *Selected geometric parameters (Å, °)*

| | | | |
|---|---|---|---|
| Ti1—Ti2 | 3.4110 (8) | Ti1—O2$^{iv}$ | 1.9586 (6) |
| Ti1—Ti2$^{i}$ | 3.5259 (8) | Ti2—O1$^{v}$ | 2.228 (2) |
| Ti1—Ti1$^{ii}$ | 3.7852 | Ti2—O1$^{iii}$ | 1.9611 (6) |
| Ti2—Ti2$^{ii}$ | 3.7852 | Ti2—O1$^{iv}$ | 1.9611 (6) |
| Ti1—Ti1$^{iii}$ | 3.1723 (7) | Ti2—O2 | 2.208 (2) |
| Ti1—Ti2$^{iii}$ | 3.1843 (7) | Ti1—Br1 | 2.5451 (7) |
| Ti2—Ti2$^{iii}$ | 3.1976 (7) | Ti1—Br2 | 2.5312 (7) |
| Ti1—O1 | 2.219 (2) | Ti2—Br1$^{v}$ | 2.5535 (7) |
| Ti1—O2 | 2.201 (2) | Ti2—Br2 | 2.5406 (7) |
| Ti1—O2$^{iii}$ | 1.9586 (6) | | |
| Ti1—Br2—Ti2 | 84.53 (2) | Ti2$^{i}$—O1—Ti2$^{iv}$ | 99.32 (6) |
| Ti1—Br1—Ti2$^{i}$ | 87.51 (2) | Ti2$^{iii}$—O1—Ti2$^{iv}$ | 149.63 (13) |
| Ti1—O2—Ti2 | 101.39 (9) | Ti1—O2—Ti1$^{iv}$ | 99.25 (6) |
| Ti1—O1—Ti2$^{i}$ | 104.91 (10) | Ti1$^{iii}$—O2—Ti1$^{iv}$ | 150.17 (13) |
| Ti1—O1—Ti2$^{iv}$ | 99.05 (6) | Ti1$^{iv}$—O2—Ti2 | 99.52 (6) |

Symmetry codes: (i) $x, y-1, z$; (ii) $x-1, y, z$; (iii) $x-\frac{1}{2}, \frac{3}{4}-y, -z$; (iv) $\frac{1}{2}+x, \frac{3}{4}-y, -z$; (v) $x, 1+y, z$.

A single crystal of TiOBr was glued onto a carbon fiber that was attached to a closed-cycle helium cryostat mounted on a four-circle Huber diffractometer. X-ray diffraction with synchrotron radiation was measured at beam-line D3 of Hasylab in Hamburg. Diffraction at room-temperature confirmed the FeOCl structure type with an orthorhombic lattice and space group Pmmn. Diffraction at 17.5 K indicated the presence of superlattice reflections at positions $q$ = (0, 1/2, 0) from the main reflections. Accordingly all Bragg reflections at low temperature were indexed with respect to the orthorhombic unit cell of the room temperature structure and the modulation wave vector $q$, employing four integers (hklm). The main reflections are described by m = 0, while m = 1 indicates the superlattice reflections.

Due to the presence of the cryostat only a limited range of setting angles of the crystal could be reached. This allowed the measurement of most Bragg reflections in one octant that represents the unique reflections with respect to *mmm* symmetry. Furthermore 16 observed main reflections and 24 observed satellite reflections could be measured within a second octant that would be required for a complete unique data set within monoclinic symmetry.

The structure was refined within the superspace formalism as a commensurately modulated structure with superspace group Pmmn(0$\beta$0) with $\beta$=1/2 and $t_0$=1/8. Section $t_0$=1/8 of superspace indicates monoclinic symmetry for the supercell. Therefore, one parameter was introduced for pseudo-merohedral twinning of the monoclinic structure on the pseudo-orthorhombic lattice. The refined volume fraction of the second twin domain was obtained as 0.492 (7). This implies a 1:1 ratio of volumes of the two twin domains, and an apparent *mmm* symmetry of the diffraction pattern. Therefore, the measured single octant represents a complete data set for the twinned crystal, as was confirmed by the good fit to the reflections measured for the second octant.

The superspace refinement allowed the structural parameters to be separated in parameters of the basic structure as defined by the strong main reflections and modulation parameters that give rise to the weak superstructure reflections (satellites). One modulation wave was refined, leading to two modulation parameters for each atom. Partial $R$-values at convergence R(main reflections) = 1.59% and R(satellites) = 4.86% show that a good fit to the superstructure reflections was obtained.

Subsequently the structure was transformed to a two-fold superstructure, leading to two independent atoms in the supercell instead of one in the basic structure unit cell. To allow for direct comparison of the superstructure and the room-temperature structure, a non-standard setting "monoclinic a unique" with inversion center at (1/4 3/8 0) was used. An ordinary structure refinement was performed in the supercell, but with additional restrictions as they were derived from the superspace approach: displacement parameters of the two independent atoms of the same species were made equal, the displacement parameter $U_{23}$ of all atoms was kept equal to 0, and $y$(A2) = (0.5+$y_{\rm ave}$-$y$(A1)), were $y_{\rm ave}$ is the $y$ coordinate of atom A in the room-temperature structure with values $y_{\rm ave}$=0.5 for $A$ = Ti and $y_{\rm ave}$ = 0 for A equal to O or Br. Omiting these restrictions leads to severe correlations between the parameters due to the pseudo-orthorhombic nature of the structure.

Data collection: *DIF*4 (Eichhorn, 1996). Cell refinement: *DIF*4 (Eichhorn, 1996). Data

reduction: Reduce (Eichhorn, 1991), Jana2000 (Petricek *et al.*, 2000). Program(s) used to refine structure: Jana2000 (Petricek *et al.*, 2000). Molecular graphics: DIAMOND - Version 2.1*c* (Brandenburg, 1999). Software used to prepare material for publication: Jana2000 (Petricek *et al.*, 2000).

Single crystals were grown by A. Suttner. Wolfgang Morgenroth is thanked for assistance with the diffraction experiments with synchrotron radiation at beam-line D3 of HASYLAB at DESY in Hamburg, Germany, and Ton Spek and Syd Hall for discussion on symmetry and space group symbols. Financial support by the German Science Foundation (DFG) is gratefully acknowledged.

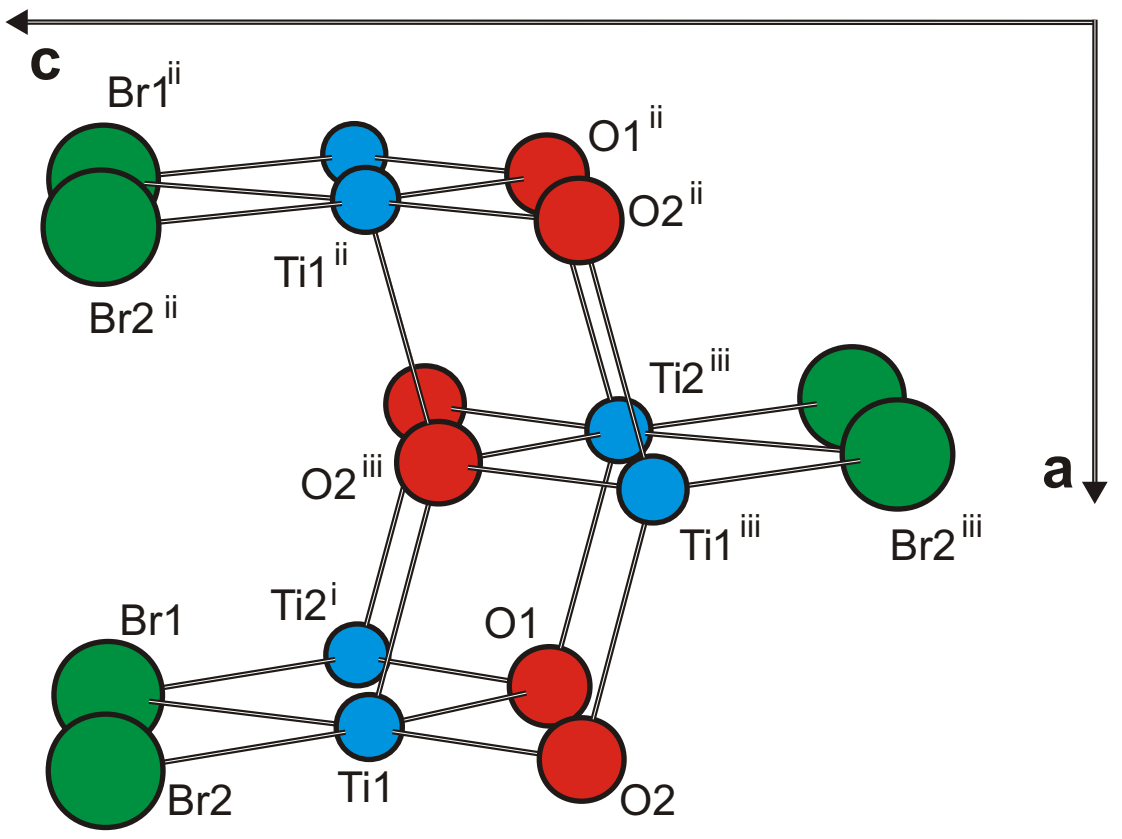

Fig. 1. Perspective view of one layer of the crystal structure of TiOBr (color online: blue for Ti, red for O, and green for Br). Atomic labels and symmetry codes correspond to those given in Table 1.

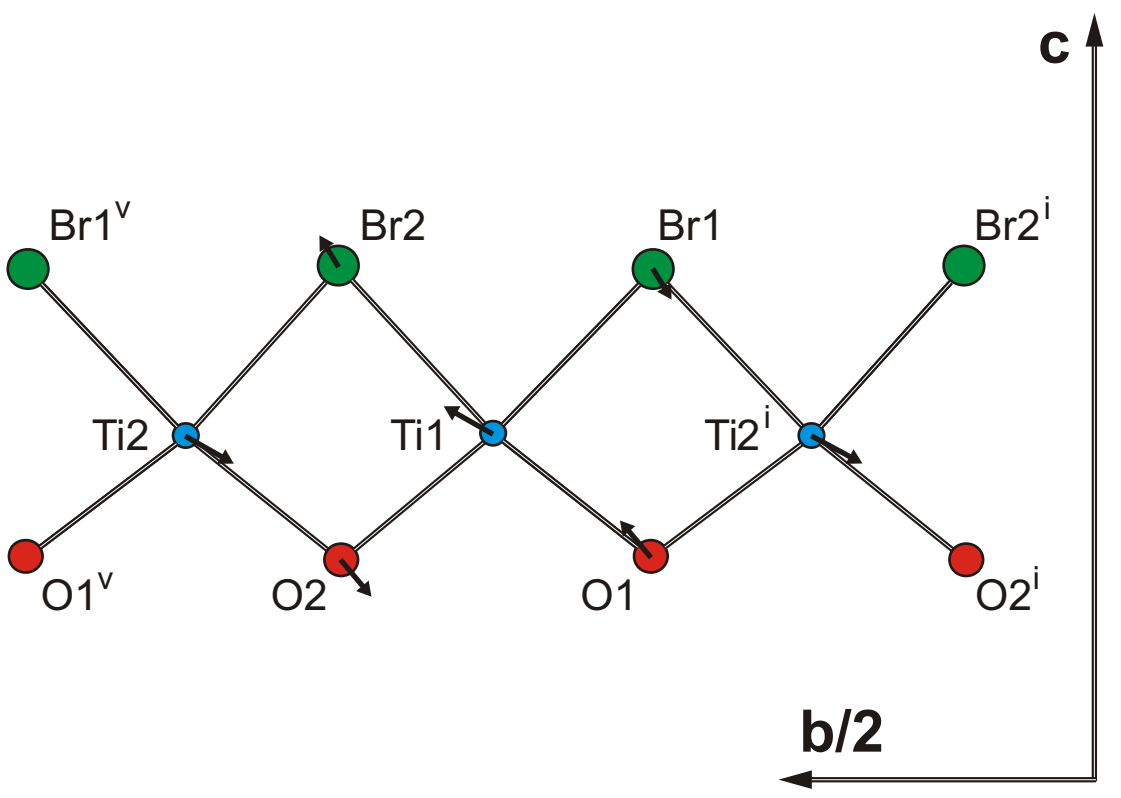

Fig. 2. One ribbon parallel [010] at x = 0, containing the chain of Ti atoms. Deviations from the average structure are given by arrows (20x their true values). Unit cell axes are indicated (color online: blue for Ti, red for O, and green for Br). Atomic labels and symmetry codes correspond to those given in Table 1.